\documentclass[aps,prd,twocolumn,superscriptaddress,amsmath,amssymb,amsthm,nofootinbib, preprintnumbers]{revtex4-2}
\usepackage{url}
\usepackage[colorlinks=true,
    linkcolor=blue,
    citecolor=blue,
    urlcolor=blue]{hyperref}
 \usepackage{amsmath}
\usepackage{graphicx}
\usepackage{epstopdf}
\usepackage{physics}
\usepackage{float}
\usepackage{hyperref}
\usepackage{color}
\usepackage[T1]{fontenc}
\usepackage[utf8]{inputenc}
\usepackage[toc,page]{appendix}
\usepackage[usenames,dvipsnames]{xcolor}
\usepackage[normalem]{ulem}

\usepackage{tikz}
\usetikzlibrary{shapes,positioning,shadows,graphs.standard,automata,arrows}
\usepackage{pgfplots}
\pgfplotsset{compat=newest, width=2.669cm, height=2.669cm, scale only axis=true,enlargelimits=false}
\pgfplotsset{tick label style={font=\tiny}}
\pgfplotsset{every major tick/.append style={major tick length=3pt}}
\pgfplotsset{every minor tick/.append style={minor tick length=1.5pt}}
\usepgfplotslibrary{groupplots}
\usepackage{caption}
\usepackage[font=small,labelfont=bf,justification=justified,format=plain]{caption}

\providecommand{\renewoperator}[3]{%
\renewcommand*{#1}{\mathop{#2}#3}}

\makeatletter
\providecommand*{\diff}%
{\@ifnextchar^{\DIfF}{\DIfF^{}}}
\def\DIfF^#1{%
\mathop{\mathrm{\mathstrut d}}%
\nolimits^{#1}\gobblespace}
\def\gobblespace{%
\futurelet\diffarg\opspace}
\def\opspace{%
\let\DiffSpace\!%
\ifx\diffarg(%
\let\DiffSpace\relax
\else
\ifx\diffarg[%
\let\DiffSpace\relax
\else
\ifx\diffarg\{%
\let\DiffSpace\relax
\fi\fi\fi\DiffSpace}

\renewoperator{\Re}{\mathrm{Re}}{\nolimits}
\renewoperator{\Im}{\mathrm{Im}}{\nolimits}

\usepackage{bm}

\newcommand{\be}{\begin{equation}}
\newcommand{\ee}{\end{equation}}
\newcommand{\ba}{\begin{eqnarray}}
\newcommand{\ea}{\end{eqnarray}}

\newcommand{\beq}{\begin{equation}}
\newcommand{\eeq}{\end{equation}}
\newcommand{\beqa}{\begin{eqnarray}}
\newcommand{\eeqa}{\end{eqnarray}}

\begin{document}

\title{Magnetized dynamical black holes}


\author{Jibril Ben Achour}
\email{j.benachour@lmu.de}
\affiliation{Arnold Sommerfeld Center for Theoretical Physics, Munich, Germany}
\affiliation{Univ de Lyon, ENS de Lyon, Laboratoire de Physique, CNRS UMR 5672, Lyon 69007, France}

\author{Adolfo Cisterna}
\email{adolfo.cisterna.r@mail.pucv.cl}
\affiliation{Sede Esmeralda, Universidad de Tarapac{\'a}, Avenida Luis Emilio Recabarren 2477, Iquique, Chile}

\author{Amaro Díaz}
\email{amdiaz2022@udec.cl}
\affiliation{Departamento de F\'isica, Universidad de Concepci\'on,
Casilla, 160-C, Concepci\'on, Chile}

\author{Keanu M{\"u}ller}
\email{keanumuller2016@udec.cl}
\affiliation{Departamento de F\'isica, Universidad de Concepci\'on,
Casilla, 160-C, Concepci\'on, Chile}


\begin{abstract}
We construct a novel exact solution of the Einstein–scalar–Maxwell equations describing a dynamical black hole immersed in an external, time-dependent electromagnetic field. Motivated by the need for more realistic analytical black hole models, our construction incorporates two key ingredients often neglected in exact solutions: a fully dynamical cosmological background and the non-perturbative backreaction of external electromagnetic fields. The compact object is obtained by dressing a Schwarzschild black hole with a radially and temporally dependent scalar field, yielding a time-dependent generalization of the Fisher–Janis–Newman–Winicour solution within the Fonarev framework. The external electromagnetic field is generated via a Lie point symmetry of the Einstein–scalar–Maxwell system, which exports the effect of a Harrison transformation to dynamical settings provided a spacelike Killing vector is present. The resulting spacetime combines a spherically symmetric dynamical horizon with an axisymmetric electromagnetic field and exhibits a rich asymptotic structure mixing Friedmann–Lemaître–Robertson–Walker and Levi–Civita geometries. We show that the time dependence of the configuration plays a crucial role in potentially cloaking curvature singularities, which would otherwise be generically naked in the stationary limit. We analyze the geometric and physical properties of the solution, including its asymptotic behavior, algebraic classification, and the structure of trapped surfaces defining the dynamical horizon. 
Possible implications for primordial black holes and some astrophysical applications, as well as extensions to higher dimensions, are also discussed.

\end{abstract}

\maketitle

\section{Introduction}

Owing to the intrinsic complexity of the Einstein field equations, the construction of exact black hole solutions generally requires a substantial level of idealization (beyond symmetry requirements), which often obscures their realistic astrophysical character. In particular, black holes are typically modelled as isolated systems, idealized as compact sources embedded in an asymptotically flat spacetime. Within this framework, interactions with surrounding fields are usually neglected, including those of significant phenomenological relevance, such as the coupling between black holes and external electromagnetic fields.
Therefore, in order to construct more realistic exact analytical black hole configurations, two key ingredients must be incorporated. First, it is desirable to understand how such local objects can be consistently embedded in dynamical spacetimes, in particular within a Friedmann–Lemaître–Robertson–Walker (FLRW) cosmological background, so that the spacetime asymptotically reproduces the FLRW behavior observed in the present Universe. Second, it is important to include the dynamical backreaction of external electromagnetic fields and to investigate how their presence modifies the structure and properties of the black hole horizon.

The study of dynamical black holes can be traced back to the seminal works of McVittie, Straus, and Tolman \cite{McVittie, Straus, Tolman}. Progress in this direction was subsequently limited for several decades, until a renewed interest emerged more recently \cite{Lasota, Thakurta,Kastor:1992nn,Brill:1993tm,Husain:1994uj,Fonarev:1994xq,Kothawala:2004fy, Gibbons:2009dr,Akcay:2010vt,Kastor:2011vz, Mello:2016irl,Kastor:2016cqs,Babichev:2018ubo, Xavier:2021chn, Croker:2021duf,Heydari:2021gea, Sato:2022yto, Babichev:2023mgk, Tang:2024cfy, Rasulian:2025jpp}. 

The construction of dynamical black hole solutions immediately encounters a significant technical difficulty. Such time-dependent spacetimes are necessarily non-vacuum, since a nontrivial field—beyond the matter content associated with the compact object—must be present to sustain the cosmological FLRW character of the background in which the localized source is embedded. This intrinsic time dependence further complicates the resolution of the Einstein equations, as a substantial degree of symmetry, most notably stationarity, is lost. In particular, the absence of an asymptotically timelike Killing vector renders many of the standard solution-generating techniques ineffective, thereby substantially increasing the complexity of the problem.  

On the other hand, the characterization of exact solutions describing delocalized external electromagnetic fields that fully backreact on the geometry presents its own set of technical challenges. First, the presence of such fields necessarily alters the asymptotic structure of the spacetime. In the most favorable scenario, the geometry becomes asymptotically conformally flat, as in the case of the Bertotti–Robinson spacetime \cite{Bertotti:1959pf,Alekseev:1996fq}. In the more extensively studied Melvin–Bonnor solution \cite{MAMelvin,Bonnor:1954tis}, however, the spacetime instead approaches a Levi–Civita–type asymptotic form, corresponding to a geometry with cylindrical symmetry. In either case, specialized solution-generating techniques are required to construct the corresponding exact spacetime \cite{Ernst:1967by,Ernst:1967wx,Harrison,Alekseev3,Alekseev1,Alekseev2}. While these mathematical methods are well established for stationary geometries, their extension to fully dynamical settings—particularly in the presence of a black hole source—remains largely unknown in exact form.

In the present work, we introduce a novel exact solution describing a Schwarzschild-like dynamical black hole immersed in an external, time-dependent electromagnetic field. The compact object is obtained by dressing a Schwarzschild black hole with a scalar field that depends on both the radial and time coordinates, yielding a time-dependent generalization of the Fisher–Janis–Newman–Winicour (FJNW) solution \cite{Fisher:1948yn,Janis:1968zz}. This construction follows the approach originally proposed by Fonarev for dynamical, spherically symmetric black holes, a method that has recently been extended to encompass arbitrary axisymmetric, albeit still non-rotating, sources \cite{BenAchour:2025vur}.
The dynamical electromagnetic field, on the other hand, is generated by exploiting the observation that the Lie point symmetry introduced in \cite{Dowker:1993bt} for Einstein–dilaton–Maxwell theory—which, in a suitable limit, applies to Einstein–scalar–Maxwell theory and reproduces at the metric level the effect of an idealized Harrison transformation \cite{Harrison}—can be employed independently of the dynamical character of the source, provided that a spacelike Killing vector is present. Since the time-dependent FJNW solution is axially symmetric, this symmetry can be readily used to magnetize the spacetime, allowing our new solution to be obtained in a straightforward manner.

Our solution consistently combines several key elements. A compact object with a spherically symmetric, dynamical horizon is embedded within an axisymmetric, time-dependent external electromagnetic field. The dynamical nature of the spacetime is supported by an appropriately chosen scalar field, whose time-dependent component drives the FLRW behavior of the geometry, while its radial dependence enables the presence of a massive compact source. The electromagnetic field, in turn, is influenced by the scalar field and induces an asymptotic structure of cylindrical Levi–Civita type.\footnote{The Levi-Civita spacetime can be represented by the line element
\begin{equation}
    ds^2=-\rho^{4\sigma}dt^2+\rho^{4\sigma(2\sigma-1)}(d\rho^2+dz^2)+\rho^{2(1-2\sigma)}d\varphi^2,
\end{equation}
where $\sigma$ represents an effective gravitational mass per unit length of a cylindrical source.  
}
The time-dependent structure of our configuration is precisely what allows, under some circumstances, the geometry to cloak the singularity associated with the massive source, which, in the presence of the scalar field, would otherwise be naked in a stationary setting. 
Unless either the time-dependent component of the scalar field or the external magnetic configuration is switched off, the asymptotic geometry generically exhibits a nontrivial interplay between FLRW and Levi–Civita behaviors.

In what follows, we present a detailed analysis of the main geometrical features of our novel spacetime configuration. Section II introduces the essential preliminaries required for the construction of the solution, including a concise overview of the Fonarev scheme and the magnetizing Lie point symmetry discussed above. In Section III, we combine these two techniques to explicitly construct the spacetime geometry and analyze its most relevant properties, with particular emphasis on the massless limit. We examine various limiting cases, the asymptotic behavior of the geometry, as well as that of the scalar and electromagnetic fields, the algebraic classification of the spacetime, and the characterization of the trapped surfaces that define the dynamical horizon. Finally, in Section IV, we discuss the physical relevance of this class of geometries and their potential applications in astrophysical processes, and we outline several directions for future research. An Appendix is included in which we present a higher-dimensional extension of our solution.

\section{Preliminaries}

To introduce our new solution, we begin by reviewing the two key ingredients underlying its construction. The first is the extended Fonarev technique \cite{Fonarev:1994xq,BenAchour:2025vur}, which allows one to uplift any axisymmetric solution of the Einstein–scalar system to a fully dynamical configuration within an augmented theory that includes a Liouville-type self-interaction for the scalar field. The second is the magnetizing Lie point symmetry \cite{Dowker:1993bt}, which makes it possible to embed such dynamical configurations consistently within external, delocalized electromagnetic fields.

\subsection{Dynamical black holes with FLRW behaviour}

Since dynamical solutions of the sort considered here are necessarily non-vacuum, it is essential to identify suitable matter fields capable of sourcing the desired cosmological behaviour. In the present case, our goal is to obtain an asymptotically FLRW spacetime, a feature that is known to be naturally supported by scalar fields. Indeed, within Einstein–scalar theory, one can readily introduce a scalar field with a logarithmic time dependence that generates the required cosmological FLRW geometry at large scales. 

The inclusion of a massive compact source within such a dynamical background is more delicate. Nevertheless, a systematic approach was developed by Fonarev \cite{Fonarev:1994xq} for the case of spherically symmetric sources, a construction that can be straightforwardly extended to the non-rotating axisymmetric setting \cite{Tangen:2007yn,BenAchour:2025vur}. 
In the most general formulation, the essential ingredient is the consideration of Einstein–scalar theory supplemented by a Liouville-type potential. The underlying theory supporting axisymmetric, non-rotating dynamical black hole solutions is therefore described by
\begin{equation}
    S=\int\sqrt{-g}d^4x\left(\frac{R}{2\kappa}-\frac{1}{2}\partial_\mu\phi\partial^\mu\phi-V(\phi)\right),
\end{equation}
with $V(\phi)$ the self-interacting potential $V(\tilde{\phi}) = V_{0} e^{\xi_3 \tilde{\phi}}$. Here, $\kappa$ represents the Newton constant and $\xi_3$ a constant parameter. 
The corresponding stationary variation of the fields provides the field equations
\begin{subequations}
\label{eomES}
\begin{align}
G_{\mu\nu} &  = \kappa \left[ \phi_{\mu} \phi_{\nu} - g_{\mu\nu} \left( \frac{1}{2} g^{\alpha\beta} \phi_{\alpha} \phi_{\beta} + V(\phi) \right)\right],\\
\Box \phi & = V_{\phi}. 
\end{align}
\end{subequations}
Given a vacuum ($\bar{R}_{\mu\nu}=0$) spacetime solution of the form 
\begin{align}
    d\bar{s}^2 =  \bar{g}_{aa}(dx^a)^2+\bar{h}_{ij}dx^idx^j, 
\end{align}
(no summation in "$a$" understood) that satisfies
 \begin{equation}
 \label{conditionsBuchdahl}
    \partial_a \bar{g}_{\mu\nu}=0= \bar{g}_{ia}, 
\end{equation}
i.e. no $a-$dependence\footnote{Therefore, "$a$" represents an ignorable coordinate along which the spacetime has no off-diagonal terms.} one can construct an $a$-dependent extension $(g, \phi)$ that solves the Einstein–Scalar system \eqref{eomES} which takes the form  
\begin{subequations}
\label{generalizedfonarev}
\begin{align}
    d s^2&=e^{2\mu(a)}[(\bar{g}_{aa})^\beta(dx_a)^2+(\bar{g}_{aa})^{1-\beta} \bar{h}_{ij}dx^idx^j],\\
    \phi&=\xi_0\ln(\bar{g}_{aa})+\frac{\xi_1}{\kappa}\mu(a),
\end{align}
\end{subequations}
where the conformal factor is given by 
\begin{equation}
\mu(a)=\xi_2\ln(Ca+T).
\end{equation}
The parameter space defined by $V_0$, $\xi_1$, $\xi_2$, and $\xi_3$ is constrained to follow 
\begin{subequations}
\label{coeff}
\begin{align}
    &\xi_{1}= - \xi_3 = \frac{\beta}{\xi_{0}} \\
    & (\beta^{2}-2\xi_{0}^{2}\kappa)\xi_{2}=2\xi_{0}^{2}\kappa  \\
    & (2\xi_{0}^{2}\kappa-\beta^{2})^{2} V_{0}=\mp \; 2\xi_{0}^{2}C^{2}(\beta^{2}-6\xi_{0}^{2}\kappa),
\end{align}
\end{subequations}
with $(-)$ if $a$ is timelike and $(+)$ if spacelike, and with constants $C$ and $T$ remaining free. Here, $\beta$ represents the so-called hair parameter that controls the appearance of the scalar field in the non-dynamical case and defines $\xi_0=\sqrt{\frac{1-\beta^2}{2\kappa}}$.

Choosing $a$ to be the time coordinate, as one would intuitively expect, allows for the construction of dynamical axisymmetric spacetimes in a straightforward manner. 
It is straightforward to observe that, in this framework, the scalar field component controlled by the conformal factor is responsible for the FLRW asymptotics, while the term governed by $\xi_0$ allows for the inclusion of a compact object. In the absence of time dependence, the incorporation of the compact source comes at the cost of introducing a curvature singularity at what would otherwise correspond to the black hole horizon. The dynamical structure of spacetime allows, under certain conditions, for this singularity to be cloaked behind a dynamical trapped surface \cite{Kastor:1992nn,Brill:1993tm,Fonarev:1994xq,Maeda:2007bu,Kastor:2011vz,BenAchour:2025vur}.

A particularly relevant case arises in the absence of self-interaction, $V_0=0$. Under this condition, the scalar hair parameter is fixed to the value $\beta=\sqrt{3}/2$.\footnote{Actually, the hair parameter is fixed to $\beta=\pm\sqrt{3}/2$, however, the negative branch is phenomenologically less interesting.} Although this scenario is less generic, it plays a central role in our analysis, since the Lie point symmetry employed to magnetize the class of dynamical axisymmetric spacetimes introduced below is applicable only when the self-interaction potential vanishes.

\subsection{Magnetizing symmetry for spacetimes with a spacelike symmetry}

In the general stationary and axisymmetric setting, the Einstein–Maxwell equations can be reformulated in terms of the Ernst formalism \cite{Ernst:1967by,Ernst:1967wx}. One of the principal advantages of this approach is that, when expressed in terms of the complex Ernst potentials, the electrovacuum field equations reveal a rich set of Lie point symmetries. These symmetries provide powerful solution-generating techniques, allowing one to construct more elaborate solutions from simple seed spacetimes while remaining within the same set of field equations.

Among these transformations, the Harrison symmetry \cite{Harrison}, belonging to the class collectively known as the Kinnersley symmetries \cite{Kinners}, acts as a charging transformation capable of introducing external electromagnetic fields into a given seed solution. A notable example is the Melvin–Bonnor spacetime \cite{MAMelvin,Bonnor:1954tis}, which can be obtained by applying a Harrison transformation to Minkowski spacetime. More generally, any stationary and axisymmetric electrovacuum solution—including black hole geometries—can serve as a suitable seed for such a transformation \cite{Ernst:1976bsr}.

Although Einstein–dilaton–Maxwell (EDM) theory does not admit an Ernst-type formulation, the authors of \cite{Dowker:1993bt} ingeniously constructed a charging symmetry that allows black hole solutions of EDM theory to be immersed in external magnetic fields. This symmetry closely resembles the Harrison transformation in its action, but is formulated directly at the level of the metric, as no potential-based reduction is available. An important feature of this construction is that, in the limit of vanishing dilaton coupling, it induces a charging mechanism for solutions of Einstein–scalar theory, irrespective of whether they are stationary. Indeed, since the symmetry is not derived from an Ernst reduction of the electrovacuum equations, it does not require stationarity and relies solely on the presence of axisymmetry. 

In this context, the symmetry acts on a scalar-vacuum seed solution $(\bar{g}_{\mu\nu},\bar{\phi})$ as follows. Owing to axisymmetry, the original metric can be decomposed as $\bar{g}_{\mu\nu}=(\bar{g}_{ij},\bar{g}_{\varphi\varphi})$. Assuming furthermore that the seed configuration is non-rotating, so that $\bar{g}_{i\varphi}=0$, the resulting charged solution takes the form (hereafter we use $\kappa=2$)
\begin{subequations}
\label{chargesymmetry}
\begin{align}
ds^2&=\Lambda^2\bar{g}_{ij}dx^idx^j+\frac{\bar{g}_{\varphi\varphi}}{\Lambda^2}d\varphi^2,\\
A&=\frac{B}{2\Lambda}\bar{g}_{\varphi\varphi}d\varphi,
\end{align}
\end{subequations}
where
\begin{equation}
\Lambda=1+\frac{B^2}{4}\bar{g}_{\varphi\varphi}.
\end{equation}
The parameter $B$ modulates the intensity of the magnetic field. The scalar field remains invariant under this transformation. It is important to emphasize, that the manner in which this magnetizing symmetry acts inevitably modifies the asymptotic structure of the original seed spacetime, replacing asymptotic flatness with an asymptotically Levi–Civita geometry characterized by a Levi–Civita parameter $\sigma$ equal to unity; see \cite{Stephani:2003tm}.
This behaviour is generic and can be understood within the Ernst framework. Indeed, any of the Kinnersley symmetries, when acting magnetically on an asymptotically flat seed, transforms the asymptotic geometry into a Levi–Civita spacetime with the parameter fixed to one. This mechanism underlies the asymptotic structure of Melvin, vortex-like, and inverted spacetimes \cite{MAMelvin,Contopoulos:2015wra,Ernst:1976bsr,Barrientos:2025rjn,Barrientos:2024pkt,Barrientos:2024uuq,DiPinto:2025yaa}.

\section{Dynamical Schwarzschild-Melvin black holes}

Having established the two mechanisms described in the previous section, we now proceed to combine them in order to construct our dynamical solution. For the sake of simplicity we start from the simplest seed, a Schwarzschild black hole.
The resulting spacetime describes a Schwarzschild-like (FJNW) dynamical black hole embedded in an external, time-dependent electromagnetic field. Using spherical-like coordinates $(t,r,\theta,\varphi)$, is given by the configuration 
\begin{widetext}
\begin{subequations}
\begin{align}
    d{s}^{2}&=(Ct+T)^{2\xi_{2}}\left\lbrace{\Lambda(t,r,\theta)^{2}\left[-f(r)^{\beta}dt^{2}+\frac{dr^{2}}{f(r)^{\beta}}+f(r)^{1-\beta}r^{2}d\theta^{2}\right]+\frac{f(r)^{1-\beta}r^{2}\sin^{2}{\theta}}{\Lambda(t,r,\theta)^{2}}d\varphi^{2}}\right\rbrace,\\
    {\phi}(t,r)&=\sqrt{\frac{1-\beta^{2}}{4}}\ln{f(r)}+\frac{\xi_{1}\xi_{2}}{2}\ln{(Ct+T)},\\
    {A}&=\frac{B}{2}\frac{(Ct+
    T)^{2\xi_{2}}f(r)^{1-\beta}r^{2}\sin^{2}{\theta}}{\Lambda(t,r,\theta)} d{\varphi},
\end{align}\label{dynamicalmagnetized}
\end{subequations}
\end{widetext}
with 
\begin{equation}
    \Lambda(t,r,\theta)=1+\frac{B^{2}}{4}(Ct+T)^{2\xi_{2}}f(r)^{1-\beta}r^{2}\sin^{2}{\theta}, 
\end{equation}
and 
\begin{equation}
f(r)=1-\frac{2m}{r}. 
\end{equation}
Although we have kept the parameters $(\xi_1,\xi_2,\beta)$ explicit in the solution, we recall that, in the absence of a self-interacting potential, they are fixed to the specific values $(\xi_1=2\sqrt{3},\xi_2=1/2,\beta=\sqrt{3}/2)$, as dictated by \eqref{coeff}. In higher dimensions, they will indeed take different values; see Appendix A. 

Note that the order in which the two techniques are combined is non-commutative. While it is possible to magnetize a dynamical solution of the Einstein–scalar system, it is not, to the best of our knowledge, possible to promote a stationary solution of Einstein–Maxwell theory to a fully dynamical one by analogous means. Then, it is natural to see how the dynamical seed via the $\varphi\varphi$-component of the metric induces a dynamical electromagnetic field.  

In the following subsections, we undertake a more detailed analysis of the principal features of our novel solution.

\subsection{An exact dynamical Melvin-Bonnor Universe}

The Melvin–Bonnor spacetime is a static, cylindrically symmetric solution of the Einstein–Maxwell equations. In its simplest form, it describes the full backreaction of a static, cylindrical magnetic field on the spacetime geometry, interpreted as a magnetic flux tube held together by its own gravitational attraction. The magnetic field lines are aligned with the symmetry axis, and their strength decays asymptotically away from the axis, thereby preventing gravitational collapse. Its line element and gauge field (purely magnetic case) are given by 
\begin{subequations}
\begin{align}
    ds^2&=\left(1+\frac{B^2}{4}\rho^2\right)^2(-dt^2+d\rho^2+dz^2)+\frac{\rho^2 d\varphi^2}{\left(1+\frac{B^2}{4}\rho^2\right)^2},\\
    A&=\frac{B}{2}\frac{\rho^2}{\left(1+\frac{B^2}{4}\rho^2\right)} d\varphi,
\end{align}
\end{subequations}
where it must be noted the use of cylindrical coordinates $\{\rho,z,\varphi\}$ under their usual interpretation. The spacetime is algebraically special—specifically of Petrov type D—and belongs to the Kundt class of solutions \cite{Stephani:2003tm}. 

As mentioned above, the Melvin–Bonnor spacetime exhibits Levi–Civita asymptotics. Building on this observation, several dynamical counterparts of the Melvin–Bonnor geometry have been constructed both in pure electrovacuum and within EDM theories \cite{Kastor:2013nha,Kastor:2015wda}. The underlying idea is that the Levi-Civita spacetime can be related, through analytic continuation of the radial distance into a timelike coordinate, to the vacuum anisotropic Kasner family of cosmological solutions. Performing such an analytic continuation onto the Melvin-Bonnor geometry, together with an appropriate change of topology, leads to the so-called Melvin cosmologies, which describe cosmological spacetimes with Kasner-type asymptotics. 
However, in these constructions, the dynamical, backreacting electromagnetic fields depend solely on the time coordinate. They are therefore purely cosmological in nature and do not directly support the presence of a well-behaved localized compact source.

It is thus natural to begin the analysis of our novel solution with the background configuration corresponding to the case $m=0$. This case describes a dynamical electromagnetic background with explicit dependence on both the time and radial coordinates. The corresponding configuration is given by
\begin{widetext}
\begin{subequations}
\label{dynamicalbackground}
\begin{align}
    d{s}^{2}&=(Ct+T)\left[{\Lambda}(t,\rho)^{2}(-dt^{2}+d\rho^{2}+dz^{2})+\frac{\rho^2d\varphi^{2}}{\Lambda(t,\rho)^{2}}\right],\\
    \phi&=\frac{\sqrt{3}}{2}\ln{(Ct+T)},\\
    A&=\frac{B}{2}\frac{\rho^{2}}{{\Lambda}(t,\rho)} (Ct+T)d\varphi,
\end{align}
\end{subequations}
\end{widetext}
where now 
\begin{align}
    {\Lambda}(t,\rho)&=1+\frac{B^{2}}{4}(Ct+T)\rho^{2}.
\end{align}

\subsubsection{Asymptotic behaviour of the fields}

We begin by analyzing the asymptotic behavior of the metric. Far from the symmetry axis, we find at leading order that
\begin{widetext}
\begin{align}\label{asymbackground}
    d{s}^{2}_{\rho\rightarrow\infty}&=(Ct+T)\left[\frac{B^2}{16}(Ct+T)^{2}\rho^4(-dt^{2}+d\rho^{2}+dz^{2})+\frac{d\varphi^{2}}{\frac{B^2}{16}(Ct+T)^2\rho^2}\right].
\end{align}
\end{widetext}
This line element represents a dynamical generalization of the Levi–Civita spacetime with parameter $\sigma=1$. Owing to its cylindrical and axial symmetry, it belongs to the broader class of Jordan–Ehlers–Kundt (JEK) geometries
\begin{equation}
    ds^2=e^{2(\gamma-\psi)}(-dt^2+d\rho^2)+e^{2\psi}dz^2+e^{-2\psi}W^2d\varphi^2,
\end{equation}
where $\gamma$, $\psi$ and $W$ are functions of $\{t,\rho\}$, and are given by 
\begin{align}
\psi=\frac{1}{2}\ln\left[\frac{B^2}{16}(Ct+T)^3\rho^4\right]=\frac{\gamma}{2},\quad W^2=(Ct+T)^2\rho^2.
\end{align}
The form of \eqref{asymbackground} may suggest that the geometry corresponds to a Kasner-type generalization of the Levi–Civita spacetime. However, there exists no choice of parameters within the Kasner family that reproduces the asymptotic behavior obtained in this limit.

On the other hand, close to the axis of symmetry $\rho\rightarrow0$ the metric behaves as 
\begin{equation}\label{axisFLRW}
    d{s}^{2}_{\rho\rightarrow 0}=(Ct+T)(-dt^{2}+d\rho^{2}+dz^{2}+\rho^2d\varphi^{2}), 
\end{equation}
consequently, the background geometry \eqref{dynamicalbackground} interpolates between a homogeneous FLRW cosmology in the vicinity of the symmetry axis and an anisotropic cosmological phase with induced Melvin-like asymptotics at large distances. Both regimes preserve cylindrical and axial symmetry. This behavior is reminiscent of the Melvin–Bonnor spacetime, which interpolates between Minkowski and Levi–Civita ($\sigma=1$) geometries while preserving boost symmetry. In contrast, owing to the dynamical nature of our background, boost symmetry is manifestly broken.

Finally, the background \eqref{dynamicalbackground} is certainly locally flat in terms of the Riemann tensor which goes as $R^{\mu\nu}{}_{\alpha\beta}\sim0$ in the limit of large $\rho$. 

The scalar field exhibits a logarithmic behavior both far from the symmetry axis and at late times. Nevertheless, its kinetic term remains finite in the limits of large $\rho$ and large $t$.

The gauge field largely inherits the characteristics of the Melvin–Bonnor magnetic field, albeit with an important distinction. Owing to the time dependence of the configuration, the magnetic field now induces a nontrivial electric component. This electric field should not be confused with the external electric field that may arise in a Melvin–Bonnor electromagnetic background; rather, it emerges solely as a consequence of the dynamical nature of our solution. 
Therefore, far and near to the symmetry axis, the magnetic potential goes as 
\begin{equation}
    A_{\rho\rightarrow\infty}\sim\frac{2}{B}d\varphi, \quad A_{\rho\rightarrow0}\sim\frac{B}{2}\rho^2d\varphi.
\end{equation}
The corresponding magnetic and electric components of the Faraday-Maxwell tensor are 
\begin{subequations}
\begin{align}
    F^{t\varphi}&=-\frac{B}{2}\frac{C}{(1+\frac{B^2}{4}(Ct+T)\rho^2)^2(Ct+T)^2},\\
    F^{\rho\varphi}&=\frac{B}{(1+\frac{B^2}{4}(Ct+T)\rho^2)(Ct+T)\rho}.
\end{align}
\end{subequations}
One observes the induction of an electric component $F^{t\varphi}$, which vanishes when $C=0$. 
The electric and magnetic fields are defined respectively as $E_{\mu}=F_{\mu\nu}u^{\nu}$ and $B_{\mu}=\frac{1}{2}\epsilon_{\mu\nu\alpha\beta}u^\nu F^{\alpha\beta}$, where $u^\mu$ denotes an appropriate comoving observer. Near the symmetry axis, the spacetime line element \eqref{axisFLRW} reduces to an FLRW geometry, allowing $u^\mu$ to be defined straightforwardly. In this region, the electric field vanishes, while the magnetic field remains constant, $B_z=B$, as in the standard Melvin–Bonnor configuration. Far from the symmetry axis, the behavior is again Melvin-like: both relevant components of the Faraday–Maxwell tensor vanish, $F^{t\varphi}=0=F^{\rho\varphi}$, and the electromagnetic field accordingly decays, as expected.

It can be readily observed that, in the asymptotic region close to the symmetry axis, the geometry exhibits a cosmological big-bang singularity (upon choosing $T=0$), while far from the axis the spacetime inherits the axial singularity characteristic of a Levi–Civita geometry with $\sigma=1$. Nevertheless, the full background described by \eqref{asymbackground} is completely regular for all values of $t$ and throughout the entire symmetry axis, as expected for a nontrivial magnetized background.

\subsubsection{Petrov type}

The algebraic properties of the Melvin–Bonnor spacetime are well established. It is algebraically special—specifically of Petrov type D—and belongs to the Kundt class \cite{Stephani:2003tm}, placing it within the non-expanding sector of the Plebański–Demiański \cite{Plebanski:1976gy} family of solutions. It admits a repeated radial, geodesic principal null direction that is expansion-free, shear-free, and twist-free. 

By contrast, our background spacetime \eqref{dynamicalbackground} is algebraically general, belonging to Petrov type I. This can be demonstrated by considering the following null tetrad
\begin{align}\label{tetradbackground}
    l^\mu&=\partial_\mu, \quad n^\mu=-\frac{1}{\Omega\Lambda}\left(\partial_\nu+\frac{\partial_r}{2}\right),\nonumber\\
    m^\mu&=\frac{1}{\sqrt{2\Omega}r\Lambda}\left(\partial_\theta+i\frac{\Lambda^2}{\sin\theta}\partial\varphi\right),\nonumber\\
    \bar{m}^\mu&=\frac{1}{\sqrt{2\Omega}r\Lambda}\left(\partial_\theta-i\frac{\Lambda^2}{\sin\theta}\partial\varphi\right), 
\end{align}
where a radial null direction is also identified, and that
satisfies, as usual, $l^\mu n_\mu=-1$ and $m_\mu \bar{m}^\mu=1$. Notice the use of Eddington-Finkelstein coordinates $\{\nu,r,\theta,\varphi\}$. 
The invariants 
\begin{align}
    I&=\Psi_0\Psi_4-4\Psi_1\Psi_3+3\Psi_2^2,\\
    J&=\det\begin{pmatrix}
\Psi_4 & \Psi_3 & \Psi_2\\
\Psi_3 & \Psi_2 & \Psi_1\\
\Psi_2&\Psi_1&\Psi_0
\end{pmatrix},
\end{align}
constructed from the self-dual Weyl tensor are shown to satisfy 
\begin{equation}
    I^3\neq27J^2. 
\end{equation}
In the static limit $(C\rightarrow0,T\rightarrow1)$, the tetrad \eqref{tetradbackground} reduces to a frame in which the Kundt character of the spacetime can be readily identified. However, it does not coincide with the tetrad aligned with the repeated principal null directions of the Weyl tensor, namely the frame that minimizes the number of non-vanishing Newman–Penrose scalars to the single nontrivial component $\Psi_2$.

\subsection{Dynamical Schwarzschild-like-Melvin 
black hole}

To analyse the general case with $m\neq0$, it is convenient to work in spherical-like coordinates $\{t,r,\theta,\varphi\}$. Most of the properties identified for the background configuration persist in the asymptotic analysis of this case, since the mass contribution enters the geometry as $\sim m/r$. The same conclusion applies to the electromagnetic and scalar fields. Along the same lines, the Petrov classification of the spacetime remains algebraically general, i.e. of type I.

A crucial qualitative difference, however, arises from the presence of the radial component of the scalar field, which induces an additional curvature singularity at $r=2m$ (besides the central singularity at $r=0$). In the non-dynamical limit—namely, in the FJNW–Melvin solution—the would-be horizon is replaced by a naked singularity. 

At this stage, the dynamical character of our novel geometry plays a central role. The spacetime that would otherwise correspond to a singular FJNW–Melvin–Bonnor configuration acquires, as a direct consequence of its time dependence, a trapped surface that effectively can play the role of a dynamical black hole horizon.

Generally, the analysis of the causal structure of dynamical spacetimes and the interpretation of their physical content are notoriously subtle and far from straightforward. This difficulty is well illustrated by some iconic examples, such as the McVittie and Thakurta geometries, whose physical status and horizon structure have only been fully clarified in relatively recent years
\cite{Nolan:1998xs, Nolan:1999wf, Nolan:1999kk, Kaloper:2010ec, Kobakhidze:2021rsh, Hutsi:2021nvs, Boehm:2021kzq, Harada:2021xze, Maciel:2024eys}.

A major complication arises from the absence of a timelike Killing vector, which renders the standard notion of Killing horizons inapplicable. Nevertheless, a robust and well-established framework for describing out-of-equilibrium dynamical horizons has been developed
\cite{Ashtekar:2025wnu, Ashtekar:2002ag, Ashtekar:2003hk, Ashtekar:2004cn, Ashtekar:2013qta}. Even within this formalism, however, the explicit identification and characterization of horizons in a given dynamical solution remain highly non-trivial. In particular, quasi-local (anti-)trapping horizons are defined in terms of the expansions of null congruences, quantities that are inherently foliation dependent \cite{Wald:1991zz, Schnetter:2005ea, Faraoni:2016xgy, Dotti:2023elh, Dotti:2025npw}. 

The case of spherically symmetric dynamical geometries is fairly understood, largely due to the existence of the Kodama vector \cite{Kodama:1979vn}. Any dynamical, spherically symmetric spacetime admits a divergence-free vector field—known as the Kodama vector—which plays a role analogous to that of a timelike Killing vector in stationary settings. This vector can be traced back to the presence of a rank-two Killing–Yano tensor, which exists in any dynamical spherically symmetric geometry \cite{Kinoshita:2024wyr}.
Among its key properties, the Kodama vector becomes null on trapping (or anti-trapping) horizons, mirroring the behavior of a Killing vector on the event horizon of a stationary black hole. More precisely, it is timelike in untrapped regions and spacelike in trapped (or anti-trapped) regions. Since the norm of the Kodama vector is foliation independent, it provides an invariant and powerful tool to characterize the causal structure of the spacetime and, in particular, to identify and analyze the dynamical horizons of spherically symmetric solutions.

Since our solution is axisymmetric rather than spherically symmetric, the use of a suitable generalization of the Kodama vector becomes necessary. Fortunately, such a generalization was introduced some time ago in \cite{Anco:2004bb}. Although it remained largely unnoticed for many years, it has recently attracted renewed interest \cite{Senovilla:2014ika, Dotti:2023elh, Dotti:2025npw} and has very recently been employed to analyze the causal structure of a novel class of uncharged, axisymmetric dynamical black holes \cite{BenAchour:2025vur}.

In the next section, we concisely introduce this generalization of the Kodama vector—known as the mean curvature vector (MCV) \cite{Anco:2004bb}—and along the lines of \cite{BenAchour:2025vur} employ it to identify and characterize the dynamical horizon of our general configuration \eqref{dynamicalmagnetized}. 

\subsubsection{Time-dependent (anti-) trapping horizons}

In order to identify the trapping horizons, we closely follow the procedure used in \cite{BenAchour:2025vur}. We refer the reader to that work for the full theoretical background behind the computation; here, we only outline the main logic.

Given a spacetime manifold $(\mathcal{M},g)$, we first introduce the induced two-dimensional metric
\begin{equation}
q_{\mu\nu}=g_{\mu\nu}+n_\mu n_\nu-s_\nu s_\mu,
\end{equation}
where the vectors $(n,s)$ satisfy the normalization and orthogonality conditions $n_\mu n^\mu=-1$, $s_\mu s^\mu=1$, and $n_\mu s^\mu=0$. The tensor $q_{\mu\nu}$ thus acts as a projector from $\mathcal{M}$ onto a two-dimensional spacelike surface $\mathcal{S}$.
The surface $\mathcal{S}$ is defined as the intersection $\mathcal{S}=\Sigma_t\cap\mathcal{B}$, where $\Sigma_t$ denotes a spacelike hypersurface at fixed time $t$, and $\mathcal{B}$ is a timelike boundary. Accordingly, $n_\mu dx^\mu$ is the unit normal to $\Sigma_t$, while $s_\mu dx^\mu$ is the unit normal to $\mathcal{B}$. In a cosmological setting, the surface $\mathcal{S}$ can be interpreted as representing the celestial sphere.

Since the surface $\mathcal{S}$ is embedded in $\mathcal{M}$, its extrinsic curvature can be decomposed into contributions associated with its embedding in $\Sigma_t$ and in $\mathcal{B}$. These are defined as
\begin{equation}
K_{\mu\nu}(n)=D_\mu n_\nu, \qquad K_{\mu\nu}(s)=D_\mu s_\nu,
\label{tracesK}
\end{equation}
where $D_\mu=q_\mu^{\ \nu}\nabla_\nu$ denotes the covariant derivative projected onto $\mathcal{S}$.

The MCV and its dual are then defined by
\begin{subequations}
\begin{align}
H^\mu\partial_\mu&=\frac{1}{2}\big(K(s)s^\mu-K(n)n^\mu\big)\partial_\mu,\\
H_{\perp}^\mu\partial_\mu&=\frac{1}{2}\big(K(s)n^\mu-K(n)s^\mu\big)\partial_\mu,
\end{align}
\end{subequations}
where $K(n)$ and $K(s)$ denote the traces of the corresponding extrinsic curvatures in \eqref{tracesK}. By construction, $H$ and $H_\perp$ are orthogonal, and their norms satisfy
\begin{equation}
H^2=-H_\perp^2=\frac{1}{2}\big(K^2(s)-K^2(n)\big).
\end{equation}

It turns out that $H_\perp$ provides a natural generalization of the Kodama vector to axisymmetric spacetimes \cite{Anco:2004bb}. Consequently, the loci at which its norm vanishes furnish an invariant characterization of trapped (or anti-trapped) surfaces. Indeed, introducing a null basis $(l_+,l_-)$ constructed from the orthonormal pair $(n,s)$ on $\mathcal{S}$, one finds that the norm of the MCV is directly related to the corresponding null expansions $\theta_\pm$ according to
\begin{equation}
|H|^2=-|H_\perp|^2=-\frac{1}{2}\theta_+\theta_- .
\end{equation}
Therefore, in order to identify the formation of a trapped horizon in our dynamical spacetime \eqref{dynamicalmagnetized}, it suffices to demonstrate the existence of loci at which the norm of the MCV vanishes.

Note that, in general, different choices of null tetrads lead to different null congruences and, hence, to different expansions, whose vanishing typically occurs on distinct, tetrad-dependent loci \cite{Faraoni:2016xgy}. Only in very special cases does the vanishing of an expansion correspond to a genuine geometric object, such as a horizon. This reflects the fact that different null congruences intersect a horizon in different ways—or may not intersect it at all—so their expansions generically display nonequivalent behavior.
A geometrically meaningful and invariant notion of expansion arises only when the null vectors are orthogonal to a fixed spacelike two-surface. This is precisely the setting of trapped and marginally trapped surfaces, as well as of quasi-local horizon frameworks. This is the situation relevant here: the null vectors used to define the MCV are orthogonal to a physically well-defined two-dimensional surface, ensuring that the associated zero-loci correctly and coordinate-independently identify trapped and anti-trapped surfaces.

Now, the two-dimensional surface $\mathcal{S}$ within our spacetime \eqref{dynamicalmagnetized} yields the induced metric $q_{\mu\nu}$ with line element 
  \begin{equation}
    ds^2_S = a^2 \left[ \Lambda^2 f^{1-\beta}r^2d\theta^2 + \frac{f^{1-\beta}r^2\sin^2{\theta}}{\Lambda^2}d\varphi^2 \right],
  \end{equation}
where we have defined $a=(Ct+T)^{\xi_2}$, and where the corresponding normal vectors are 
\begin{equation}
   n_\mu dx^\mu = -a\Lambda f^{\frac{\beta}{2}}dt, \quad s_\mu dx^\mu = a \Lambda f^{\frac{-\beta}{2}}dr,
 \end{equation}
which, in turn, provide the extrinsic curvature traces 
\begin{subequations}
    \begin{align}
        K(n)&= 2 \frac{f^{-\beta/2}\dot{a}}{\Lambda^2 a^2} \\
        K(s)&= f^{\beta/2}\frac{\left[2-\frac{(\beta-1)f'}{f}\right]}{r\Lambda a }. 
    \end{align}
\end{subequations}
With these ingredients in place, we evaluate the norm of the MCV
\begin{widetext}
\begin{align}
  H^\mu H_\mu = \frac{1}{f^\beta \Lambda^2 a^2} \left[ \frac{\left((\beta+1)M-r\right)^2f^{2\beta-2}}{r^4}  - \mathcal{H}^2\right], 
\end{align}
\end{widetext}
which, as discussed above, delineates the boundary between trapped and untrapped regions. Here, as usual, $\mathcal{H}=\dot{a}/a$ denotes the Hubble parameter. This expression for the norm correctly reproduces the case discussed in \cite{BenAchour:2025vur} for $\delta=1$ when $B=0$, which corresponds precisely to the norm of the MCV for the dynamical FJNW solution originally introduced in \cite{Husain:1994uj}. 

At this stage, the existence of a boundary separating trapped and untrapped regions can already be seen. In fact, upon substituting the corresponding parameter values of our solution space, $\xi_2=1/2$ and $\beta=\sqrt{3}/2$, the loci at which this interface forms—namely, the loci at which $H^2=0$ can be numerically observed; see Fig.1.\footnote{For $\beta=-\sqrt{3}/2$, the S-curve simply shows one cosmological horizon, which is the reason why we do not further analyze this case \cite{Husain:1994uj,Faraoni:2015ula}.}
\begin{figure}[h]
    \centering
    \includegraphics[width=1\linewidth]{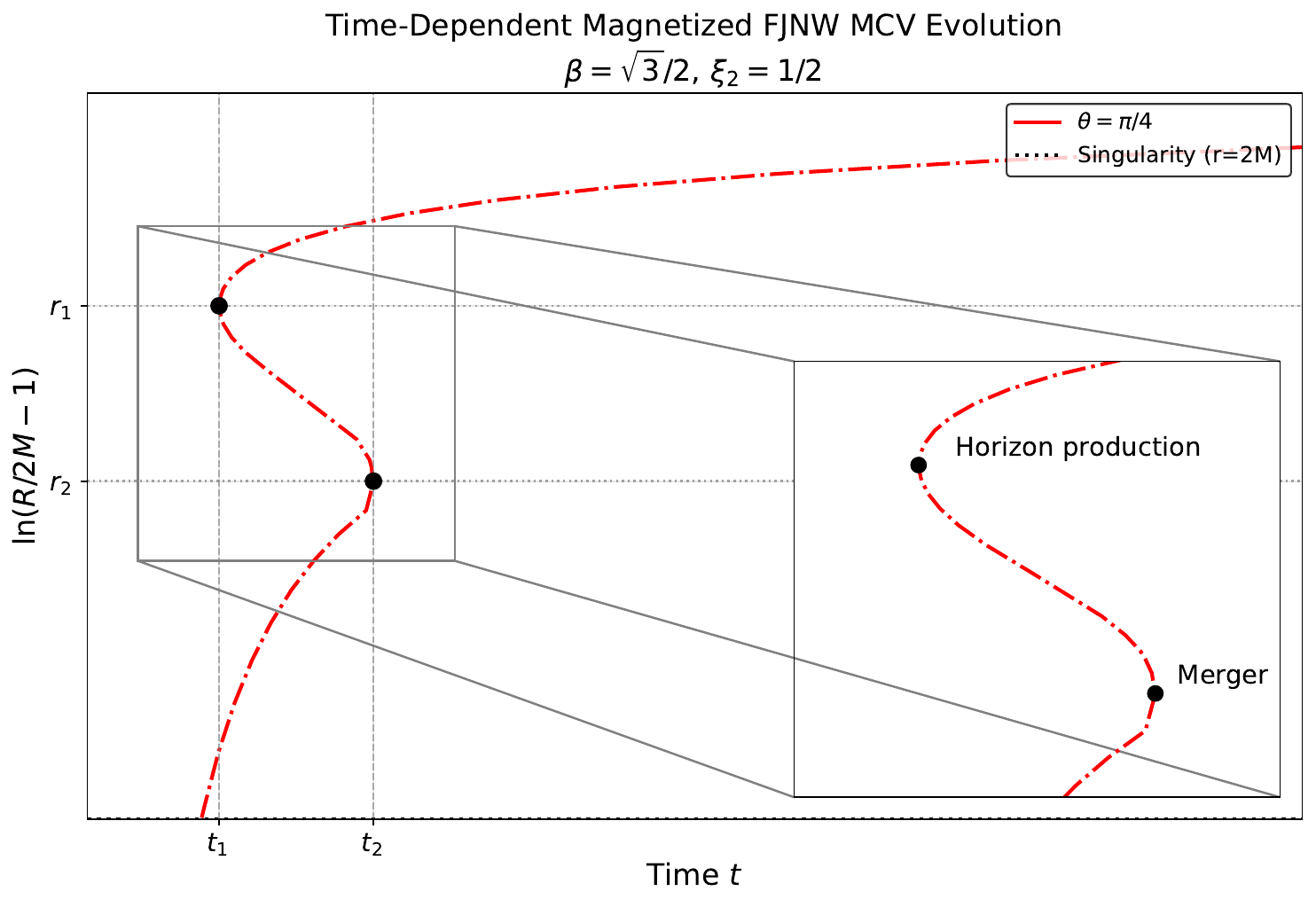}
    \caption{This is the so-called S-curve associated with the contour of $H^\mu H_\mu = 0$. Several values of $\theta$ can be considered with little change in the curve. Two sets of critical points, $(t_1,r_1)$ and $(t_2,r_2)$ are highlighted.} 
    \label{fig:placeholder}
\end{figure}

There, we observe a curve typically associated with solutions in which no self-interaction is present \cite{Husain:1994uj,BenAchour:2025vur}. Two pairs of critical points, $(t_1,r_1)$ and $(t_2,r_2)$, are highlighted. The radial coordinate $r$ has been rescaled by $2m$, so that the outer curvature singularity is located at $r=1$, which marks the innermost boundary of the spacetime.

The physical interpretation of this curve is as follows. From the initial big-bang singularity at $t=0$ up to the first critical time $t_1$, the spacetime contains a single horizon, namely a cosmological horizon that expands over time. Just after $t_1$, two additional horizons appear. One of them is again a cosmological horizon that expands indefinitely, while the other is a contracting horizon associated with the compact object. This latter horizon shrinks over time until it merges with the original cosmological horizon at the second critical time $t_2$. For $t>t_2$, only the newly formed cosmological horizon remains, which is ultimately a spacetime with a naked singularity at $r=2m$ embedded in a FLRW cosmology.

Consequently, the physically most relevant region of the spacetime seems to be the one defined by $t\in(t_1,t_2)$ and $r>r_2$. In this domain, the geometry describes a dynamical black hole embedded in a cosmological background, with a compact horizon shielding the curvature singularity at $r=2m$ and a non-compact cosmological horizon at large scales.\footnote{However, this interpretation should be approached with caution, as it remains open to debate. Due to the genuinely dynamical nature of the solution, no eternal event horizon forms; consequently, the spacetime does not contain a black hole in the global sense. At most, one can say that a temporary dynamical horizon forms and covers the singularity for a finite period, representing only a transient trapped region. From this perspective, the entire solution is more appropriately viewed as a cosmological spacetime containing a central singularity whose ability to trap light is temporary rather than permanent.}

At this stage, we loosely refer to the contracting horizon as a black hole horizon. However, it could, in principle, also correspond to a white hole horizon. To determine its true physical nature, we proceed as shown in \cite{BenAchour:2025vur}. 

We begin by constructing the null basis $(l_\pm)$, which takes the form\footnote{With respect to this basis, the induced metric on $\mathcal{S}$ is $q_{\mu\nu}=g_{\mu\nu}+l_\mu^+ l_{\nu}^-+l_\mu^- l_{\nu}^+$.}
\begin{subequations}
\begin{align}
l_+&=(n^\mu+s^\mu)\partial_\mu=\frac{1}{a\Lambda}\left(f^{-\beta/2}\partial_t+f^{\beta/2}\partial_r\right),\\
l_-&=\frac{1}{2}(n^\mu-s^\mu)\partial_\mu=\frac{1}{2a\Lambda}\left(f^{-\beta/2}\partial_t-f^{\beta/2}\partial_r\right).
\end{align}
\end{subequations}
The corresponding expansions, defined as $\theta_{\pm}=\frac{1}{\sqrt{\sigma}}\mathcal{L}_{l{\pm}}\sqrt{\sigma}$, where $\sigma$ denotes the determinant of the induced two-dimensional metric $q_{\mu\nu}$, are given by
\begin{equation}
\theta_{\pm}=\frac{\sqrt{2}}{f^{\beta/2}a\Lambda}\left[\mathcal{H}\pm \frac{f^{\beta-1}\bigl(r-m(\beta+1)\bigr)}{r^2}\right].
\end{equation}
These expressions consistently reproduce the norm of the mean curvature vector, in agreement with the relation $H_\mu H^\mu=-\tfrac{1}{2}\theta_+\theta_-$ \cite{BenAchour:2025vur}. 

To analyze the nature of the possible horizons, we first note that the solution admits two distinct branches: one with $C>0$ and $0<t<\infty$, and another with $C<0$ and $-\infty<t<0$. In what follows, we analyze both branches independently, closely following the analysis provided in \cite{BenAchour:2025vur,Faraoni:2015ula}. 

We start by focusing on the negative branch defined by $C<0$ and $-\infty<t<0$. A future outer trapping horizon corresponding to a black hole horizon is identified by the conditions (with $l_+$ and $l_-$ chosen to be future directed)

\begin{equation}
\theta_+=0,\qquad \theta_-<0,\qquad \mathcal{L}_{l_-}\theta_+<0. \label{BHCON}
\end{equation}

Setting $\theta_+=0$ leads to the relationship
\begin{equation}
\frac{\xi_2}{t_h}=-\mathcal{R}_{(r_h,\theta_h)}, \qquad
\text{with}\quad
\mathcal{R}=\frac{f^{\beta-1}\bigl(r-m(\beta+1)\bigr)}{r^2},
\label{negstivehorizonbranch}
\end{equation}
where we have explicitly substituted the Hubble factor and evaluated the expression at the horizon location $(t_h,r_h,\theta_h)$. This expression sets the horizon curve.
Since $t<0$, one has $(Ct)^{\xi_2}>0$, and because $f(r)>0$ for all $r>2m$, the sign of $\theta_-$ is controlled solely by the sign of $\xi_2$. In our solution $\xi_2=1/2>0$,\footnote{Note that this value does not change in the case $\beta\rightarrow-\beta$, and thus its sign is always fixed.} which implies $\theta_-<0$, explicitly

\begin{equation}
\theta_-=\frac{2\sqrt{2}}{f_h^{\beta/2}\Lambda_h a_h}\frac{\xi_2}{t_h}<0.
\end{equation}
Furthermore, since $t<0$ and $\xi_2>0$, the condition \eqref{negstivehorizonbranch} admits solutions only for $\mathcal{R}>0$. 

The last inequality in \eqref{BHCON} plays a crucial role in understanding the phenomenology encoded in the S-curve shown in Fig. 1 and the ultimate nature of the horizon curve. Explicitly, we have 
\begin{widetext}
\begin{align}
    \mathcal{L}_{l_-} \theta_+ = \frac{\sqrt{2}}{\Lambda^2 a^2 f^\beta} \left\{ \dot{\mathcal{H}} - f^\beta \mathcal{R}' + \left[f^\beta \partial_r \left( \frac{\beta}{2}  \ln{f} + \ln{\Lambda}\right)  - \partial_t(\ln{\Lambda}) - \mathcal{H} \right] (\mathcal{H}+\mathcal{R}) \right\} |_{(t_h,r_h,\theta_h)}.
\end{align}
\end{widetext}
In particular, the contracting event horizon is confined to the radial interval
\begin{equation}
r\in\left[r_1=\sqrt{3}+1+\frac{\sqrt{2}}{2},r_2=\sqrt{3}+1-\frac{\sqrt{2}}{2}\right],\label{values}
\end{equation} 
within which the condition $\mathcal{L}_{l_-}\theta_+<0$ is satisfied.
Within this range of the radial coordinate, the horizon is of black hole type. Outside this interval, however, the Lie derivative changes sign, $\mathcal{L}_{l_-}\theta_+>0$, signaling a qualitative change in the nature of the horizon. This sign flip marks the appearance of the cosmological horizons: the one emerging at $r_1$, which expands indefinitely, and the initially present one that eventually merges with the contracting black hole horizon at $r_2$. Consequently, the loci at which $\mathcal{L}_{l_-}\theta_+$ changes sign define the critical points $(t_1,t_2)$ and $(r_1,r_2)$, which govern the transitions between black hole and cosmological horizon behaviour.

With this in mind, we can now determine the nature of the cosmological background supporting the dynamical compact object. To this end, we examine the behaviour of the Hubble factor
\begin{equation}
\mathcal{H}=\frac{\xi_2}{t}<0, \qquad \mathcal{\dot{H}}+\mathcal{H}^2=\frac{\xi_2(\xi_2-1)}{t^2}<0.
\end{equation}
These relations show that the spacetime corresponds to a contracting cosmological background that is decelerating. Consequently, the compact object endowed with a future black hole trapping horizon is embedded in a contracting Universe \cite{Kastor:1992nn,Brill:1993tm,Husain:1994uj,Fonarev:1994xq,Faraoni:2015ula}. It is also worth noting that past anti-trapping horizons—characterized by the conditions $\theta_-=0$ and $\theta_+>0$—cannot arise in the present setup, since their existence would require $\xi_2<0$, which is excluded for our solution.

Now, let us conclude this analysis with the dissection of the branch $C>0$ and $0<t<\infty$. 
In this case $t>0$. $a(t)=(Ct)^{\xi_2}>0$ and again $f(r)>0$ for any $r>2m$. Due to these conditions, we are concerned with the appearance of a past outer anti-trapping horizon governed by 
\begin{equation}
\theta_-=0,\qquad \theta_+>0,\qquad \mathcal{L}_{l_+}\theta_-<0. \label{WHCON}
\end{equation}
Proceeding in analogy with the previous case, we get from $\theta_-=0$ that
\begin{equation}
    \frac{\xi_2}{t_h}=\mathcal{R}_{(r_h,\theta_h)}, 
\end{equation}
relation that now provides the horizon curve.
We observe that 
\begin{equation}
    \theta_+=\frac{2\sqrt{2}}{f_h^{\beta/2}\Lambda_h a_h}\frac{\xi_2}{t_h}>0,
\end{equation}
is always positive as $\xi_2=1/2$. In addition, 
\begin{widetext}
\begin{align}
    \mathcal{L}_{l_+} \theta_- = \frac{\sqrt{2}}{2\Lambda^2 a^2 f^\beta} \left\{ \dot{\mathcal{H}} - f^\beta \mathcal{R}' - \left[f^\beta \partial_r \left( \frac{\beta}{2}  \ln{f} + \ln{\Lambda}\right)  + \partial_t(\ln{\Lambda}) + \mathcal{H} \right] (\mathcal{H}-\mathcal{R}) \right\}|_{(t_h,r_h,\theta_h)}.
\end{align}
\end{widetext}
Again, the inequality associated with this last expression determines the nature of our horizon curve. The last inequality in \eqref{WHCON} is satisfied for the same range $r\in(r_1,r_2)$. In that context, the horizon is represented by a past outer anti-trapping horizon associated with a white hole, while outside of it, the cosmological horizons take place. 

Finally, from the Hubble function we observe that 
\begin{equation}
\mathcal{H}=\frac{\xi_2}{t}>0, \qquad \mathcal{\dot{H}}+\mathcal{H}^2=\frac{\xi_2(\xi_2-1)}{t^2}<0.
\end{equation}
Thus, this situation corresponds to a past outer anti-trapping horizon given by a white hole immersed in an expanding, decelerating cosmology.

\section{Discussion}

We have constructed an exact dynamical spacetime that, among its causal structures, describes a dynamical black hole immersed in an external, time-dependent electromagnetic field. This result opens two complementary avenues of investigation: a primarily theoretical one, of interest to the community working on exact solutions in GR, and a more phenomenological one, related to the physics of dynamical black holes and their potential astrophysical signatures.

On the theoretical side, we have shown that it is possible to consistently combine the Fonarev construction \cite{Fonarev:1994xq} for generating dynamical solutions in Einstein–scalar theory with standard magnetization techniques in GR \cite{Harrison,Ernst:1976bsr,Dowker:1993bt}. While the scope and robustness of the Fonarev scheme are now well understood \cite{BenAchour:2025vur}, its interplay with magnetizing symmetries, to the authors knowledge, remained unexplored. 

The underlying reason behind the existence of our solution is clear: although the lack of stationarity—or, more generally, the absence of two commuting Killing vectors—precludes an Ernst reduction of the field equations \cite{Ernst:1967by,Ernst:1967wx}, Lie point symmetries of the type employed here only require the emergence of a nonlinear sigma model after reduction along a single Killing vector, whether timelike or spacelike. In our case, this vector is provided by the axial symmetry of the spacetime, and the resulting sigma model naturally accommodates a minimally coupled scalar field.

An important feature of our solution \eqref{dynamicalmagnetized} is that it is not merely time-dependent, but also exhibits a nontrivial radial dependence. This sharply distinguishes it from previous constructions involving time-dependent electromagnetic backgrounds, which are often purely cosmological. Moreover, our spacetime contains a genuine compact source and is not restricted to the background limit $m=0$. The combined backreaction of the scalar and electromagnetic fields endows the geometry with a rich and unconventional asymptotic structure. A more detailed investigation of these time-dependent electromagnetic configurations—particularly their possible radiative character and the existence of additional symmetries—would be highly desirable.

From a phenomenological perspective, our results suggest new directions for exploration. Astrophysical black holes are widely believed to power relativistic plasma jets extending over enormous distances, observed in systems ranging from supermassive black holes in active galactic nuclei to stellar-mass black holes in X-ray binaries. Understanding the mechanisms responsible for launching and sustaining these jets, as well as for particle acceleration, remains a long-standing challenge. Among the existing models, the Blandford–Znajek \cite{Blandford:1977ds} mechanism plays a central role: it provides an exact solution of the force-free Maxwell equations on the Kerr background and predicts a magnetosphere with a nontrivial Poynting flux that extracts energy from a rotating black hole. It is now well established that such electromagnetic energy fluxes can be sustained by stationary, spinning black hole magnetospheres.

The solution presented here points to a qualitatively different possibility. Although the black hole is non-rotating, its dynamical magnetosphere exhibits a nonvanishing electromagnetic energy flux, encoded in the $T_{\rho\varphi}$ component of the electromagnetic energy–momentum tensor. Remarkably, this flux is directly proportional to the parameter $C$, which controls the time dependence of the spacetime. The main obstacle to a detailed physical interpretation lies in the fact that the geometry is neither asymptotically flat nor purely asymptotically FLRW. Nevertheless, to the best of our knowledge, this constitutes the first exact solution describing a non-rotating, magnetized black hole with a nonvanishing Poynting flux, opening new perspectives on energy extraction mechanisms driven by spacetime dynamics rather than rotation alone.

\section{Acknowledgments} 
The authors gratefully acknowledge insightful discussions with José Barrientos and Professor Valerio Faraoni. 
A.C. is partially supported by FONDECYT grant 1250318. The work of K.M. is funded by Beca Nacional de Doctorado ANID grant No. 21231943.
Keanu Müller would like to dedicate this work to the memory of Bladimir Figueroa for his lifelong commitment to the education and mentorship of students.
\bigskip
\section{Appendix: Higher dimensional construction}

As is well known, the Fonarev construction is rooted in the existence of the Buchdahl theorem \cite{Buchdahl:1959nk}, which holds in arbitrary spacetime dimensions. As a consequence, the extension of the Fonarev technique to higher dimensions follows in a straightforward manner. On the other hand, the magnetizing symmetry employed in this work has also been formulated in arbitrary dimensions \cite{Ortaggio:2004kr}. Taken together, these results make the construction of higher-dimensional dynamical black holes immersed in dynamical magnetic fields conceptually direct. The higher-dimensional generalization of our configuration \eqref{dynamicalmagnetized} is therefore given by
\begin{widetext}
\begin{subequations}
\label{FinalSolutionDdimensional}
\begin{align}
    ds^{2}&=(Ct+T)^{2\xi_{2}}\qty{\Lambda(t,r,\theta)^{\frac{2}{D-3}}\qty[-f(r)^{\beta}dt^{2}+\frac{dr^{2}}{f(r)^{\beta}}+f(r)^{\frac{1-\beta}{D-3}}r^{2}\qty(d\theta^{2}+\cos^{2}{\theta}d\Omega^{2}_{D-4})]+\frac{f(r)^{\frac{1-\beta}{D-3}}r^{2}\sin^{2}{\theta}}{\Lambda(t,r,\theta)^{2}}d\varphi^{2}},\\
    \phi(t,r,\theta)&=\xi_{0}\log{f(r)}+\xi_{1}\xi_{2}\frac{(D-2)}{4}\log{(Ct+T)},\\
    A(t,r,\theta)&=\frac{B}{2}\frac{(Ct+T)^{2\xi_{2}}f(r)^{\frac{1-\beta}{D-3}}r^{2}\sin^{2}{\theta}}{\Lambda(t,r,\theta)}d\varphi,
\end{align}
\end{subequations}
\end{widetext}
with
\begin{align}
    \Lambda(t,r,\theta)&=1+\frac{B^{2}}{2}\frac{(D-3)}{(D-2)}(Ct+T)^{2\xi_{2}}f(r)^{\frac{1-\beta}{D-3}}r^{2}\sin^{2}{\theta},
\end{align}
and 
\begin{equation}
    f(r)=1-\frac{\mu}{r^{D-3}}. 
\end{equation}
Here, $\theta\in\qty[0,\pi/2]$, $\varphi\in[0,2\pi)$, and $d\Omega^{2}_{D-4}=d\phi_{1}^{2}+\sin^{2}{\phi_{1}}d\phi_{2}^{2}+...+\prod_{i=1}^{D-5}\sin^{2}{\phi_{i}}d\phi^{2}_{D-4}$. In addition, $\mu>0$ is a parameter proportional to the physical mass.
Furthermore, since the transformation assumes a configuration with a vanishing potential term in the action, the parameter $\beta$ must be fixed as follows
\begin{align}
    \beta&=\frac{\sqrt{2(D-1)(D-2)}}{2(D-2)}.
\end{align}
Consequently, the remaining parameters are given by
\begin{align}
 \xi_0=\frac{1}{4},\quad  \xi_{1}=\frac{2\sqrt{2(D-1)(D-2)}}{(D-2)},\quad\xi_{2}&=\frac{1}{(D-2)}.
\end{align}
Finally, \eqref{FinalSolutionDdimensional} constitutes a configuration that satisfies the Einstein-Maxwell-scalar field equations in higher dimensions.

\newpage

\bibliography{apssamp}

\end{document}